\documentclass[twocolumn,eqsecnum,aps]{revtex4}
\usepackage[dvips]{epsfig,color}
\usepackage{graphicx}
\usepackage{amsmath,amsfonts,amsbsy,amssymb}

\begin{document}

\title{Grassmannian Fields and Doubly Enhanced Skyrmions\\
in Bilayer Quantum Hall system at $\nu=2$}
\author{K. Hasebe and Z.F. Ezawa}
\affiliation{Department of Physics, Tohoku University, Sendai, 980-8578 Japan }

\begin{abstract}
\vspace*{5mm} At the filling factor $\nu=2$ the bilayer quantum Hall system
has three phases, the ferromagnetic phase (spin phase), the spin singlet
phase (ppin phase) and the canted antiferromagnetic phase. We analyze soft
waves and quasiparticle excitations in the spin and ppin phases. It is shown
that the dynamic field is the Grassmannian G$_{4,2}$ field carrying four
complex degrees of freedom. In each phase there are four complex soft waves
(pseudo-Goldstone modes) and one kind of skyrmion excitations (G$_{4,2}$
skyrmions) flipping either spins or pseudospins coherently. An intriguing
property is that a quasiparticle is a G$_{4,2}$ skyrmion essentially
consisting of two CP$^{3}$ skyrmions and thus possesses charge $2e$.
\end{abstract}

\maketitle

\section{Introduction}

Quantum Hall (QH) effects have attracted renewed attention\cite%
{BookDasSarma,BookEzawa} owing to its peculiar features associated with
quantum coherence. Skyrmions\cite{Sondhi93B} are charged excitations
reversing several electron spins coherently, whose existence has been
established firmly in monolayer QH systems\cite%
{Barrett95L,Aifer96L,Schmeller95L}. Bilayer quantum Hall (BLQH) systems are
much more interesting because they exhibit unique QH effects originating in
the interlayer interaction. For instance, we may have not only skyrmions in
the spin space but also skyrmions in the layer (pseudospin) space\cite%
{Ezawa97B,Ezawa99L}. Furthermore, an anomalous tunneling current has been
observed\cite{Spielman00L} between the two layers at the zero bias voltage.
It may well be a manifestation of the Josephson-like phenomena predicted a
decade ago\cite{Ezawa92IJMPB}. They occur due to quantum coherence developed
spontaneously across the layers\cite{Ezawa97B,Moon95B}.

Electrons make cyclotron motions and their energies are quantized into
Landau levels under perpendicular magnetic field $B_{\perp }$. The number
density of magnetic flux quanta is $\rho _{\Phi }\equiv B_{\perp }/\Phi _{%
\text{D}}$, where $\Phi _{\text{D}}=2\pi \hbar /e$ is the flux unit. One
electron occupies an area $2\pi \ell _{B}^{2}$ with $\ell _{B}=\sqrt{\hbar
/eB_{\perp }}$ the magnetic length. The filling factor is $\nu =\rho
_{0}/\rho _{\Phi }$ with $\rho _{0}$ the electron number density. In the
BLQH system one Landau site may accommodate four isospin states $|\text{f}%
\!\uparrow \rangle $, $|\text{f}\!\downarrow \rangle $, $|\text{b}%
\!\!\uparrow \rangle $, $|\text{b}\!\!\downarrow \rangle $ in the lowest
Landau level (LLL), where $|\text{f}\!\uparrow \rangle $ implies that the
electron is in the front layer and its spin is up and so on. The SU(4)
symmetry underlies the BLQH system provided the cyclotron energy is large
enough.

The driving force of quantum coherence is the Coulomb exchange interaction.
It is described by an anisotropic SU(4) nonlinear sigma model in BLQH
systems, as is most easily demonstrated in the von-Neumann-lattice
formulation\cite{EzawaX02B}. The BLQH system at $\nu =1$ has been
extensively analyzed\cite{EzawaX02B,RJ} based on this SU(4) scheme, where
the dynamic field is the CP$^{3}$ field carrying three complex degrees of
freedom. It has been shown at $\nu =1$ that there are three complex soft
waves (pseudo-Goldstone modes) and one kind of skyrmion excitations (CP$^{3}$
skyrmions) flipping both spins and pseudospins coherently.

New features appear in the BLQH system at $\nu =2$, where two
distinguishable phases have clearly been observed experimentally\cite%
{Pelle97,Sawada98L,Sawada99R}. One is a fully spin polarized ferromagnetic
phase, which we call the spin phase. The other is a spin singlet phase,
which we call the pseudospin (abridged as ppin) phase because it is also a
fully pseudospin polarized one. (The layer degree of freedom is referred to
as the pseudospin.) It has been argued that there arises a new phase, called
canted antiferromagnetic phase\cite{Demler99L,Zheng97L,Sarma97L}, between
the spin and ppin phases in the phase diagram. It is realized by the effect
of the SU(4)-noninvariant part of the Coulomb exchange interaction. However,
according to an exact numerical diagonalization method on few-electron
systems\cite{Schliemann00L} the canted phase occupies a very tiny region in
the phase diagram and is practically negligible, as is consistent with
experimental results\cite{Pelle97,Sawada98L,Sawada99R}.

The aim of this paper is to analyze soft waves and topological solitons in
the spin and ppin phases. The Hilbert space in the lowest Landau level\ is
spanned by the six states occupying two of the four isospin states. 
An intriguing property is that one quasiparticle consists of two skyrmions
and thus possesses charge $2e$, as was pointed out in Ref.\cite%
{BookEzawa,Ezawa99L}. However, the previous analysis is primitive and
incomplete. We present a theory of the $\nu =2$ BLQH system based on the
SU(4) scheme with the Grassmannian manifold. We show that the dynamic field
is the Grassmannian G$_{4,2}$ field carrying four complex degrees of freedom
and that the Coulomb exchange interaction is described by the Grassmannian G$%
_{4,2}$ sigma model. In each phase there are four complex soft waves
(pseudo-Goldstone modes) and one kind of skyrmion excitations (G$_{4,2}$
skyrmions) flipping either spins or pseudospins coherently. We confirm that
one G$_{4,2}$ skyrmion essentially consists of two CP$^{3}$ skyrmions and
carries charge $2e$.

In section \ref{SecLandaHamil}, after reviewing the microscopic Landau-site
Hamiltonian\cite{EzawaX02B} for the BLQH system, we present an improved
expression for the Coulomb exchange interaction. In section \ref%
{SecGrounState} we make a group theoretical study of isospin states in the
BLQH system. At $\nu =2$ they belong to the $\boldsymbol{6}$-dimensional
representation of SU(4). Classifying them with respect to the subgroup SU$_{%
\text{spin}}$(2)$\otimes $SU$_{\text{ppin}}$(2), we introduce the spin phase
and the ppin phase. In section \ref{SecSU4Break}, we investigate the SU(4)
invariant limit of the $\nu =2$ BLQH system, where the symmetry SU(4) is
spontaneously broken into U(1)$\otimes $SU(2)$\otimes $SU(2). There are
eight broken generators; $S_{x}$, $S_{y}$, $2S_{x}P_{a}$ and $2S_{y}P_{a}$
in the spin phase, and $P_{y}$, $P_{z}$, $2S_{a}P_{y}$ and $2S_{a}P_{z}$ in
the ppin phase, where $S_{a}$ and $P_{a}$ are respectively the generators of
the groups SU$_{\text{spin}}$(2) and SU$_{\text{ppin}}$(2). Accordingly
there arise eight Goldstone modes in the SU(4) invariant limit. The target
space becomes the Grassmannian G$_{4,2}$ manifold, whose real 
dimension is eight.
In section \ref{SecGrass}, we introduce the composite-boson (CB) picture of
electrons to make a further study of the Goldstone modes. The CP$^{3}$ field
is defined as the normalized CB field. To describe two electrons in one
Landau site we introduce two CP$^{3}$ fields. By requiring that two
electrons are indistinguishable, the set of two CP$^{3}$ fields turns out to
be the Grassmannian G$_{4,2}$ field $Z(\boldsymbol{x})$. It is a $4\times 2$
matrix field carrying eight real independent components representing the
Goldstone modes. In section \ref{SecSoftWave}, we construct the Hamiltonian
in the $\boldsymbol{6}$-dimensional representation of SU(4), where the basic
field is the Grassmannian field $Z(\boldsymbol{x})$. We analyze the soft modes
which are the pseudo-Goldstone modes made gapful by the Zeeman or tunneling
gap. In section \ref{SecGrassSolit} we analyze topological solitons (G$%
_{4,2} $ skyrmions), whose existence follows from the homotopy theorem, $\pi
_{2}($G$_{4,2})=\mathbb{Z}$, with $\mathbb{Z}$ the integer additive group.
Analyzing the condition that they are confined within the lowest Landau
level, we find that they are comprised of two skyrmions excited in the front
and back layers for the spin phase, or in the up-spin and down-spin states
for the ppin phase. We call them biskyrmions. In section \ref{SecEffecTheor}%
, we present an effective spin-1 theory of biskyrmions. It is shown that
they are represented as the well-known O(3) skyrmions in this effective
spin-1 theory. In section \ref{SecTwoMono} we study the criterion whether
the system is to be regarded as a genuine BLQH system with biskyrmion
excitations or as a set of two monolayer QH systems with simple skyrmion
excitations. Recent experimental data\cite{Kumada00L} have revealed a
remarkable difference in the activation energy behavior between two bilayer
samples with small and large tunneling gaps. We explain it based on this
criterion.

\section{Exchange Interactions\label{SecLandaHamil}}

The kinetic Hamiltonian of planar electrons in the bilayer system is given
by 
\begin{equation}
H_{\text{K}}={\frac{1}{2M}}\int d^{2}x\boldsymbol{\psi }^{\dagger }(\boldsymbol{x}%
)(D_{x}-iD_{y})(D_{x}+iD_{y})\boldsymbol{\psi }(\boldsymbol{x})  \label{HamilKinem}
\end{equation}%
apart from the cyclotron energy $N\hbar \omega _{\text{c}}/2$, where $%
D_{k}=-i\hbar \partial _{k}+eA_{k}$. The electron field $\boldsymbol{\psi }$
possesses the SU(4) isospin index $\sigma =\text{f}\!\uparrow ,\text{f}%
\!\downarrow ,\text{b}\!\uparrow ,\text{b}\!\downarrow $. When the cyclotron
gap is large enough, thermal excitations across Landau levels are
practically impossible. Hence, it is a good approximation to neglect all
those excitations by requiring the confinement of electrons to the lowest
Landau level. This leads to the LLL condition on the state, 
\begin{equation}
(D_{x}+iD_{y})\psi ^{\sigma }(\boldsymbol{x})|\varphi \rangle =0,
\label{LLLcondi}
\end{equation}%
implying that the kinetic energy (\ref{HamilKinem}) vanishes. It determines
the Hilbert space $\mathbb{H}_{\text{LLL}}$.

We analyze electrons confined to the lowest Landau level. One Landau site
contains four electron states distinguished by the SU(4) isospin index $%
\sigma $. The group SU(4) is generated by the Hermitian, traceless, $4\times
4$ matrices. There are $(4^{2}-1)$ independent matrices. We take a standard
basis\cite{BookGeorgi}, $\lambda _{a}$, $a=1,2,\cdots ,15$, normalized as $%
\text{Tr}(\lambda _{a}\lambda _{b})=2\delta _{ab}$. They are the
generalization of the Pauli matrices.

The Coulomb interaction is given by 
\begin{equation}
H_{C}=\sum_{\alpha ,\beta =\text{f},\text{b}}{\frac{1}{2}}\int
d^{2}xd^{2}yV_{\alpha \beta }(\boldsymbol{x}-\boldsymbol{y})\rho _{\alpha }(\boldsymbol{x%
})\rho _{\beta }(\boldsymbol{y}),
\end{equation}%
where $V_{\text{f}\text{f}}=V_{\text{b}\text{b}}={e^{2}}/4\pi \varepsilon r$
is the intralayer Coulomb interaction, while $V_{\text{f}\text{b}}=V_{\text{b%
}\text{f}}={e^{2}}/4\pi \varepsilon \sqrt{r^{2}+d^{2}}$ is the interlayer
Coulomb interaction with $d$ the interlayer separation. The Coulomb
interaction is decomposed into two terms, $H_{C}=H_{C}^{+}+H_{C}^{-}$, with 
\begin{subequations}
\label{BLCouloPM}
\begin{align}
H_{C}^{+}& ={\frac{1}{2}}\int d^{2}xd^{2}yV_{+}(\boldsymbol{x}-\boldsymbol{y})\rho (%
\boldsymbol{x})\rho (\boldsymbol{y}),  \label{BLCouloP} \\
H_{C}^{-}& ={\frac{1}{2}}\int d^{2}xd^{2}yV_{-}(\boldsymbol{x}-\boldsymbol{y})\Delta
\rho (\boldsymbol{x})\Delta \rho (\boldsymbol{y}),  \label{BLCouloM}
\end{align}%
where $V_{\pm }={\frac{1}{2}}(V_{\text{f}\text{f}}\pm V_{\text{f}\text{b}})$%
; $H_{C}^{+}$ depends on the total density $\rho (\boldsymbol{x})$, while $%
H_{C}^{-}$ on the density difference $\Delta \rho (\boldsymbol{x})$ between the
front and back layers, 
\end{subequations}
\begin{equation}
\Delta \rho (\boldsymbol{x})=\rho ^{\text{f}\uparrow }(\boldsymbol{x})+\rho ^{\text{f%
}\downarrow }(\boldsymbol{x})-\rho ^{\text{b}\uparrow }(\boldsymbol{x})-\rho ^{\text{%
b}\downarrow }(\boldsymbol{x}).
\end{equation}%
The Coulomb term $H_{C}^{+}$, which is invariant under the SU(4)
transformation, dominates the BLQH system provided the interlayer separation 
$d$ is small enough.

The Coulomb exchange interaction is the key to quantum coherence. An easiest
way for its derivation is to expand the electron field $\psi ^{\sigma }(%
\boldsymbol{x})$ as 
\begin{equation}
\psi ^{\sigma }(\boldsymbol{x})=\sum_{i=1}^{N_{\Phi }}c_{\sigma }(i)\varphi _{i}(%
\boldsymbol{x}),  \label{FieldExpanBL}
\end{equation}%
where $N_{\Phi }$ is the total number of Landau sites; $c_{\sigma }(i)$ is
the annihilation operator of the electron with isospin $\sigma $ at Landau
site $i$, 
\begin{align}
\{c_{\sigma }(i),c_{\tau }^{\dagger }(j)\}& =\delta _{ij}\delta _{\sigma
\tau },  \notag \\
\{c_{\sigma }(i),c_{\tau }(j)\}& =\{c_{\sigma }^{\dagger }(i),c_{\tau
}^{\dagger }(j)\}=0,  \label{AntiCommuC}
\end{align}%
and $\varphi _{i}(\boldsymbol{x})$ is the one body wave function determined to
satisfy the LLL condition (\ref{LLLcondi}). It describe an electron
localized around the Landau site $i$.

We substitute the expansion (\ref{FieldExpanBL}) into the Coulomb
interaction terms (\ref{BLCouloP}) and (\ref{BLCouloM}). From the
SU(4)-invariant term (\ref{BLCouloP}), we obtain a Landau-site Hamiltonian%
\cite{EzawaX02B} representing the SU(4)-invariant exchange interaction, 
\begin{equation}
H_{X}^{+}=-4\sum_{\langle i,j\rangle }J_{ij}^{+}\biggl(\boldsymbol{T}(i)\!\cdot
\!\boldsymbol{T}(j)+{\frac{1}{8}}n(i)n(j)\biggr).  \label{BCouloIVBL}
\end{equation}%
Here, the sum $\sum_{\langle i,j\rangle }$ runs over spin pairs just once;$\ 
\boldsymbol{T}=(T_{1},T_{2},\ldots ,T_{15})$ is the SU(4) isospin, 
\begin{equation}
T_{a}=(c_{\text{f}\uparrow }^{\dagger },c_{\text{f}\downarrow }^{\dagger
},c_{\text{b}\uparrow }^{\dagger },c_{\text{b}\downarrow }^{\dagger }){\frac{%
\lambda _{a}}{2}}%
\begin{pmatrix}
c_{\text{f}\uparrow } \\ 
c_{\text{f}\downarrow } \\ 
c_{\text{b}\uparrow } \\ 
c_{\text{b}\downarrow }%
\end{pmatrix}%
,  \label{SinLayer}
\end{equation}%
and $n(i)$ is the electron number operator, $n\equiv \sum_{\sigma }c_{\sigma
}^{\dagger }c_{\sigma }$, at each site $i$. The exchange integral $%
J_{ij}^{\pm }$ is given by 
\begin{equation}
J_{ij}^{\pm }={\frac{1}{2}}\int \!d^{2}xd^{2}y\;\varphi _{i}^{\ast }(\boldsymbol{%
x})\varphi _{j}^{\ast }(\boldsymbol{y})V_{\pm }(\boldsymbol{x}-\boldsymbol{y})\varphi
_{i}(\boldsymbol{y})\varphi _{j}(\boldsymbol{x}).
\end{equation}%
It is notable that the exchange term (\ref{BCouloIVBL}) is rewritten as 
\begin{align}
H_{X}^{+}=-8\sum_{\langle i,j\rangle }& J_{ij}^{+}\bigl(\boldsymbol{S}(i)\!\cdot
\!\boldsymbol{S}(j)+{\frac{1}{4}}n(i)n(j)\bigr)  \notag \\
& \check{\otimes}\bigl(\boldsymbol{P}(i)\!\cdot \!\boldsymbol{P}(j)+{\frac{1}{4}}%
n(i)n(j)\bigr),  \label{inv}
\end{align}%
where $\boldsymbol{S}=\boldsymbol{S}^{\text{f}}+\boldsymbol{S}^{\text{b}}$ and $\boldsymbol{P%
}=\boldsymbol{P}^{\uparrow }+\boldsymbol{P}^{\downarrow }$. The symbol $\check{%
\otimes}$ is understood as the direct product with respect to the Pauli
matrices such as 
\begin{equation}
S_{a}P_{b}\equiv S_{a}\check{\otimes}P_{b}=(c_{\text{f}\uparrow }^{\dagger
},c_{\text{f}\downarrow }^{\dagger },c_{\text{b}\uparrow }^{\dagger },c_{%
\text{b}\downarrow }^{\dagger }){\frac{\tau _{a}}{2}}\otimes {\frac{\tau _{b}%
}{2}}%
\begin{pmatrix}
c_{\text{f}\uparrow } \\ 
c_{\text{f}\downarrow } \\ 
c_{\text{b}\uparrow } \\ 
c_{\text{b}\downarrow }%
\end{pmatrix}%
.
\end{equation}%
Various SU(2) operators are given by 
\begin{align}
S_{a}^{\text{f}}& =(c_{\text{f}\uparrow }^{\dagger },c_{\text{f}\downarrow
}^{\dagger }){\frac{\tau _{a}}{2}}%
\begin{pmatrix}
c_{\text{f}\uparrow } \\ 
c_{\text{f}\downarrow }%
\end{pmatrix}%
, & \;S_{a}^{\text{b}}& =(c_{\text{b}\uparrow }^{\dagger },c_{\text{b}%
\downarrow }^{\dagger }){\frac{\tau _{a}}{2}}%
\begin{pmatrix}
c_{\text{b}\uparrow } \\ 
c_{\text{b}\downarrow }%
\end{pmatrix}%
,  \notag \\
P_{a}^{\uparrow }& =(c_{\text{f}\uparrow }^{\dagger },c_{\text{b}\uparrow
}^{\dagger }){\frac{\tau _{a}}{2}}%
\begin{pmatrix}
c_{\text{f}\uparrow } \\ 
c_{\text{b}\uparrow }%
\end{pmatrix}%
, & \;P_{a}^{\downarrow }& =(c_{\text{f}\downarrow }^{\dagger },c_{\text{b}%
\downarrow }^{\dagger }){\frac{\tau _{a}}{2}}%
\begin{pmatrix}
c_{\text{f}\downarrow } \\ 
c_{\text{b}\downarrow }%
\end{pmatrix}%
.  \label{ISN}
\end{align}%
The equivalence between the Hamiltonians (\ref{BCouloIVBL}) and (\ref{inv})
is seen by expanding the latter as%
\begin{align}
H_{X}^{+}=-& 2\sum_{\langle i,j\rangle }J_{ij}^{+}\biggl(\boldsymbol{S}%
(i)\!\cdot \!\boldsymbol{S}(j)+\boldsymbol{P}(i)\!\cdot \!\boldsymbol{P}(j)  \notag \\
& +4S_{a}P_{b}(i)\!\cdot \!S_{a}P_{b}(j)+{\frac{1}{4}}n(i)n(j)\biggr).
\label{Exx}
\end{align}%
We may take $S_{a}$, $P_{a}$, $2S_{a}P_{b}$ instead of $T_{a}$ as the
fifteen generators of the group SU(4). These are related by 
\begin{align}
& S_{x}=T_{1}+T_{13},\quad S_{y}=T_{2}+T_{14},  \notag \\
& \qquad S_{z}=T_{3}-(1/\sqrt{3})T_{8}+(\sqrt{6}/3)T_{15},  \notag \\
& P_{x}=T_{4}+T_{11},\quad P_{y}=T_{5}+T_{12},  \notag \\
& \qquad P_{z}=(2/\sqrt{3})T_{8}+(2/\sqrt{6})T_{15},  \notag \\
& 2S_{x}P_{x}=T_{6}+T_{9},\quad 2S_{x}P_{y}=T_{7}+T_{10},  \notag \\
& \qquad 2S_{x}P_{z}=T_{1}-T_{13},  \notag \\
& 2S_{y}P_{x}=-T_{7}+T_{10},\quad 2S_{y}P_{y}=T_{6}-T_{9},  \notag \\
& \qquad 2S_{y}P_{z}=T_{2}-T_{14},  \notag \\
& 2S_{z}P_{x}=T_{4}-T_{11},\quad 2S_{z}P_{y}=T_{5}-T_{12},  \notag \\
& \qquad 2S_{z}P_{z}=T_{3}+(1/\sqrt{3})T_{8}-(\sqrt{6}/3)T_{15}.
\label{T2SP}
\end{align}%
Note that $2\boldsymbol{T}^{2}=\boldsymbol{S}^{2}+\boldsymbol{P}^{2}+(2S_{a}P_{b})^{2}$.

The exchange energy due to the SU(4)-noninvariant term (\ref{BLCouloM}) is
also evaluated. Combining them we obtain 
\begin{align}
H_{X}=& -8\sum_{\langle i,j\rangle }J_{ij}\biggl(\boldsymbol{S}(i)\!\cdot \!%
\boldsymbol{S}(j)+\frac{1}{4}n(i)n(j)\biggr)  \notag \\
& \check{\otimes}\biggl(\sum_{a}^{x,y,z}\frac{J_{ij}^{a}}{J_{ij}}%
P_{a}(i)P_{a}(j)+\frac{1}{4}n(i)n(j)\biggr),  \label{Ex}
\end{align}%
where $J_{ij}^{x}=J_{ij}^{y}=J_{ij}^{d}\equiv J_{ij}^{+}-J_{ij}^{-}$ and $%
J_{ij}^{z}=J_{ij}\equiv J_{ij}^{+}+J_{ij}^{-}$. The exchange term represents
interactions between isospins in two sites due to the overlapping of their
wave functions. For instance, the term $\boldsymbol{S}\!\cdot \!\boldsymbol{S}$
contains 
\begin{equation}
\bigl\{\boldsymbol{S}(i)\!\cdot \!\boldsymbol{S}(j)\bigr\}_{xy}\equiv {\frac{1}{2}}%
\bigl(S_{+}(i)S_{-}(j)+S_{-}(i)S_{+}(j)\bigr),  \label{ladd}
\end{equation}%
where ${S}_{\pm }=S_{x}\pm iS_{y}$ are the ladder operators. The term (\ref%
{ladd}) has a role to flip the up-spin in $j$-site and the down-spin in $i$%
-site simultaneously and vise versa. Thus, the exchange term is the origin
of isospin modulation.

\section{Ground States}

\label{SecGrounState}

We classify isospin states at $\nu =2$, where one Landau site contains two
electrons. Each electron belongs to the $\boldsymbol{4}$-dimensional irreducible
representation of SU(4). A pair of electrons is 
%Two electrons are attached to one magnetic flux and combined to form 
%a boson. It is identified with a Schwinger boson\cite{Sachdev90B} 
classified according to the group-theoretical composition rule, 
\begin{equation}
\boldsymbol{4}\otimes \boldsymbol{4}=\boldsymbol{1}\boldsymbol{0}\oplus \boldsymbol{6}.
\label{44to106}
\end{equation}%
The $\boldsymbol{1}\boldsymbol{0}$-dimensional irreducible representation is a
symmetric state, while the $\boldsymbol{6}$-dimensional irreducible
representation is an antisymmetric state. Two electrons in one Landau site
must form an antisymmetric state due to the Pauli exclusion principle.
Hence, the allowed representation is the antisymmetric $\boldsymbol{6}$%
-dimensional irreducible representation.

In the language of the subgroup SU$_{\text{spin}}$(2)$\otimes $SU$_{\text{%
ppin}}$(2), the $\boldsymbol{6}$-dimensional irreducible representation of the
group SU(4) is divided into two different irreducible representations, 
\begin{equation}
\boldsymbol{6}=(\boldsymbol{3},\boldsymbol{1})+(\boldsymbol{1},\boldsymbol{3}),
\end{equation}%
where $\boldsymbol{3}$ is the symmetric representation of SU(2), and $\boldsymbol{1}$
is the antisymmetric representation of SU(2). Consequently, there are 
%the Schwinger boson takes 
six states at each Landau site, which are grouped into two sectors, i.e.,
the $(\boldsymbol{3},\boldsymbol{1})$ sector and the $(\boldsymbol{1},\boldsymbol{3})$
sector. We call the $(\boldsymbol{3},\boldsymbol{1})$ sector the spin sector and the 
$(\boldsymbol{1},\boldsymbol{3})$ sector the ppin sector.

\begin{figure}[tbph]
\includegraphics*[width=75mm]{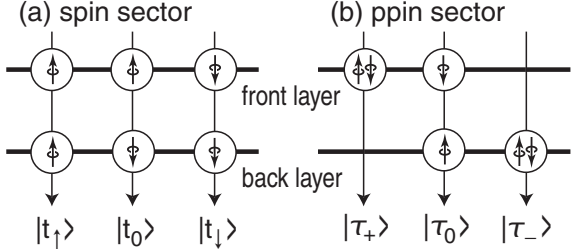}
\caption{ (a) The spin sector is comprised of spin-triplet ppin-singlet
states, $|t_{\uparrow}\rangle$, $|t_{0}\rangle$ and $|t_{\downarrow}\rangle$%
. (b) The ppin sector is comprised of spin-singlet ppin-triplet states, $|%
\protect\tau_{+}\rangle$, $|\protect\tau_{0}\rangle$ and $|\protect\tau%
_{-}\rangle$. }
\label{SectorSP}
\end{figure}

The spin sector $(\boldsymbol{3},\boldsymbol{1})$ consists of spin-triplet
pseudospin-singlet states [Fig.\ref{SectorSP}(a)], 
\begin{align}
& |t_{\uparrow }\rangle =|\text{f}^{\uparrow },\text{b}^{\uparrow }\rangle
,\quad \quad |t_{0}\rangle ={\frac{1}{\sqrt{2}}}\bigl(|\text{f}^{\uparrow },%
\text{b}^{\downarrow }\rangle +|\text{f}^{\downarrow },\text{b}^{\uparrow
}\rangle \bigr),  \notag \\
& |t_{\downarrow }\rangle =|\text{f}^{\downarrow },\text{b}^{\downarrow
}\rangle ,  \label{GrounT}
\end{align}%
where $|\text{f}^{\uparrow },\text{b}^{\uparrow }\rangle \equiv c_{\text{f}%
\uparrow }^{\dagger }c_{\text{b}\uparrow }^{\dagger }|0\rangle $, etc.

The ppin sector $(\boldsymbol{1},\boldsymbol{3})$ consists of spin-singlet
pseudospin-triplet states [Fig.\ref{SectorSP}(b)], 
\begin{align}
& |\tau _{+}\rangle =|\text{f}^{\uparrow },\text{f}^{\downarrow }\rangle
,\quad \quad |\tau _{0}\rangle =\frac{1}{\sqrt{2}}\bigl(|\text{f}^{\uparrow
},\text{b}^{\downarrow }\rangle -|\text{f}^{\downarrow },\text{b}^{\uparrow
}\rangle \bigr),  \notag \\
& |\tau _{-}\rangle =|\text{b}^{\uparrow },\text{b}^{\downarrow }\rangle ,
\label{GrounTau}
\end{align}%
where $|\text{f}^{\uparrow },\text{f}^{\downarrow }\rangle \equiv c_{\text{f}%
\uparrow }^{\dagger }c_{\text{f}\downarrow }^{\dagger }|0\rangle $, etc.

In the SU(4)-invariant limit all these six states are degenerate. Actually
the degeneracy is resolved by various SU(4)-noninvariant interactions. We
write down direct interaction terms to determine the ground state as well as
perturbative excitations. Because the QH system is robust against density
fluctuations, the direct Coulomb term arising from the SU(4)-invariant term (%
\ref{BLCouloP}) is irrelevant as far as perturbative fluctuations are
concerned. The SU(4)-noninvariant term (\ref{BLCouloM}) yields the
capacitance term, 
\begin{equation}
H_{\text{cap}}=\varepsilon _{\text{cap}}\sum_{i=1}^{N_{\Phi
}}P_{z}(i)P_{z}(i),  \label{CapacEnerg}
\end{equation}%
where $P_{z}=P_{z}^{\uparrow }+P_{z}^{\downarrow }$ at each site and 
\begin{equation}
\varepsilon _{\text{cap}}=\frac{e^{2}}{4\pi \varepsilon \ell _{B}}\sqrt{%
\frac{\pi }{2}}\biggl(1-e^{d^{2}/2\ell _{B}^{2}}\bigl\{1-\text{erf}\bigl(d/%
\sqrt{2}\ell _{B})\bigr\}\biggr)  \label{veps}
\end{equation}%
with the error function $\text{erf}(x)$. The term $\varepsilon _{\text{cap}%
}P_{z}(i)P_{z}(i)$ represents the capacitance energy per one Landau site.
Including the Zeeman and tunneling terms, the direct interaction is
summarized as 
\begin{equation}
H_{D}=\sum_{i=1}^{N_{\Phi }}\biggl(-{\Delta }_{\text{Z}}S_{z}(i)+{{%
\varepsilon }_{\text{cap}}}P_{z}(i)P_{z}(i)-{\Delta }_{\text{SAS}}P_{x}(i)%
\biggr),  \label{BDct}
\end{equation}%
where ${\Delta }_{\text{Z}}$ and ${\Delta }_{\text{SAS}}$ are the Zeeman and
tunneling gaps.

The total Landau-site Hamiltonian is 
\begin{equation}
H_{\text{total}}=H_{X}+H_{D},  \label{TotalHamil}
\end{equation}%
which is the sum of the exchange term (\ref{Ex}) and the direct term (\ref%
{BDct}).

The ground state and its energy are obtained by diagonalizing the
Hamiltonian (\ref{TotalHamil}). The Hilbert space consists of six states.
Because the spin and ppin sectors belong to different irreducible
representations of SU$_{\text{spin}}$(2)$\otimes $SU$_{\text{ppin}}$(2), we
have $\langle t_{a}|\boldsymbol{S}|\tau _{b}\rangle =0$ and so on. It follows
that $\langle t_{a}|H_{D}|\tau _{b}\rangle =0$ for any $a$ and $b$.
Therefore, the spin and ppin sectors are decoupled completely as far as the
direct interaction is concerned. This is the case also for the
SU(4)-invariant part of the Coulomb exchange interaction. These two sectors
are mixed only by the SU(4)-noninvariant exchange interaction. The mixing
between the spin and ppin sectors is absent in the vanishing limit of the
interlayer separation ($d\rightarrow 0$). It is reasonable to start with
this limit and then improve approximation. Thus, we diagonalize the
Hamiltonian in the spin and ppin sectors to obtain the ground state. The
SU(4)-noninvariant part of the Coulomb exchange interaction is included into
the ground-state energy by way of its expectation value.

In the spin sector, the energies are given by 
\begin{align}
E_{t_{\uparrow }}& =-2J-\Delta _{Z},  \notag \\
E_{t_{0}}& =-2J,  \notag \\
E_{t_{\downarrow }}& =-2J+\Delta _{Z}.  \label{TrueEnergS}
\end{align}%
The ground state of the spin sector is $\prod_{i=1}^{N_{\Phi }}|t_{\uparrow
}\rangle _{i}$ with the energy $N_{\Phi }E_{t_{\uparrow }}$.

In the ppin sector, the direct interaction reads 
\begin{equation}
H_{D}=\sum_{i}^{N_{\Phi }}%
\begin{pmatrix}
\varepsilon _{\text{cap}} & -\frac{1}{\sqrt{2}}{\Delta }_{\text{SAS}} & 0 \\ 
-\frac{1}{\sqrt{2}}{\Delta }_{\text{SAS}} & 0 & \ -\frac{1}{\sqrt{2}}{\Delta 
}_{\text{SAS}} \\ 
0 & -\frac{1}{\sqrt{2}}{\Delta }_{\text{SAS}} & \varepsilon _{\text{cap}}%
\end{pmatrix}%
_{i}.
\end{equation}%
At each site the eigenstates are given by 
\begin{align}
|v_{+}\rangle & =\frac{\cos \theta -\sin \theta }{2}(|\tau _{+}\rangle
+|\tau _{-}\rangle )+\frac{\cos \theta +\sin \theta }{\sqrt{2}}|\tau
_{0}\rangle ,  \notag  \label{TrueGrounP} \\
|v_{0}\rangle & =\frac{1}{\sqrt{2}}(|\tau _{+}\rangle -|\tau _{-}\rangle ), 
\notag \\
|v_{-}\rangle & =\frac{\cos \theta +\sin \theta }{2}(|\tau _{+}\rangle
+|\tau _{-}\rangle )+\frac{\cos \theta -\sin \theta }{\sqrt{2}}|\tau
_{0}\rangle ,
\end{align}%
where 
\begin{equation}
\tan \theta =\frac{\varepsilon _{\text{cap}}}{2{\Delta }_{\text{SAS}}+\sqrt{4%
{\Delta }_{\text{SAS}}^{2}+\varepsilon _{\text{cap}}^{2}}},
\end{equation}%
with the eigenenergies 
\begin{align}
E_{v_{+}}& =-(J+J^{d}\cos ^{2}(2\theta ))+{\frac{1}{2}}\bigl(\varepsilon _{%
\text{cap}}-\sqrt{4{\Delta }_{\text{SAS}}^{2}+\varepsilon _{\text{cap}}^{2}}%
\bigr),  \notag \\
E_{v_{0}}& =-(J+J^{d}\cos ^{2}(2\theta ))+\varepsilon _{\text{cap}},  \notag
\\
E_{v_{-}}& =-(J+J^{d}\cos ^{2}(2\theta ))+{\frac{1}{2}}\bigl(\varepsilon _{%
\text{cap}}+\sqrt{4{\Delta }_{\text{SAS}}^{2}+\varepsilon _{\text{cap}}^{2}}%
\bigr),  \label{TrueEnergP}
\end{align}%
respectively. The ground state of the ppin sector is $\prod_{i=1}^{N_{\Phi
}}|v_{+}\rangle _{i}$ with the energy $N_{\Phi }E_{v_{+}}$.

\begin{figure}[htbp]
\includegraphics*[width=75mm]{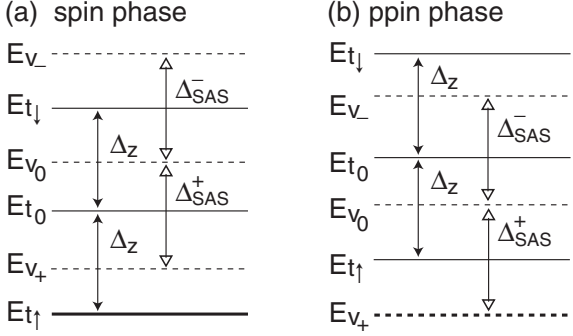}
\caption{The energy levels are comprised of six levels. The spin phase or
the ppin phase is realized for $E_{t_{\uparrow}} <E_{v_+}$ or $E_{v_+}
<E_{t_{\uparrow}}$. Here, ${\Delta}_{\text{SAS}}^{\pm}={\frac{1}{2}}\left(%
\protect\sqrt{4{\Delta }_{\text{SAS}}^{2}+\protect\varepsilon_{\text{cap}%
}^{2}}\pm\protect\varepsilon _{\text{cap}}\right)$}
\label{EneLevSP}
\end{figure}

Consequently, there are two possible ground states, $\prod_{i=1}^{N_{\Phi
}}|t_{\uparrow }\rangle _{i}$ or $\prod_{i=1}^{N_{\Phi }}|v_{+}\rangle _{i}$%
. When $E_{t_{\uparrow }}<E_{v_{+}}$, the ground state is $%
\prod_{i=1}^{N_{\Phi }}|t_{\uparrow }\rangle _{i}$, which we call the spin
phase since all spins are polarized. On the other hand, when $%
E_{v_{+}}<E_{t_{\uparrow }}$, the ground state is $\prod_{i=1}^{N_{\Phi
}}|v_{+}\rangle _{i}$, which we call the ppin phase [Fig \ref{EneLevSP}].

\section{SU(4) Breaking and Topological Solitons}

\label{SecSU4Break}

Isospin states in the $\nu =2$ BLQH system belong to the $\boldsymbol{6}$%
-dimensional irreducible representation of SU(4) as in (\ref{44to106}). We
restrict the Hamiltonian to this representation. In the SU(4)-invariant
limit the Landau-site Hamiltonian is 
\begin{equation}
H_{X}^{+}=-4\sum_{\langle i,j\rangle }J_{ij}^{+}\biggl(\widehat{\boldsymbol{T}}%
(i)\!\cdot \!\widehat{\boldsymbol{T}}(j)+{\frac{1}{8}}n(i)n(j)\biggr),
\label{SU4Site}
\end{equation}%
where $\widehat{\boldsymbol{T}}(j)$ are the fifteen generators of SU(4) in this
representation. They are the spin-1 operators. We derive a field-theoretical
Hamiltonian by taking a continuum limit. This can be done straightforwardly%
\cite{EzawaX02B} when Landau sites are taken on lattice points of a von
Neumann lattice\cite{vonNeumann55,Perelomov71,Bargmann71,Tao86}. It yields
the SU(4) nonlinear sigma model, 
\begin{equation}
{\mathcal{H}}_{X}^{+}=2J^{+}\partial _{k}\widehat{\boldsymbol{T}}(\boldsymbol{x}%
)\!\cdot \!\partial _{k}\widehat{\boldsymbol{T}}(\boldsymbol{x}),  \label{SU4Sigma}
\end{equation}%
where $J^{+}=\frac{1}{2}(J+J^{d})$ with%
\begin{equation}
J=\frac{1}{16\sqrt{2\pi }}\frac{e^{2}}{4\pi \varepsilon \ell _{B}},
\label{SpinStiff}
\end{equation}%
and%
\begin{equation}
J^{d}=J\biggl(-\sqrt{\frac{2}{\pi }}\frac{d}{\ell _{B}}+\bigl(1+\frac{d^{2}}{%
\ell _{B}^{2}}\bigr)e^{d^{2}/2\ell _{B}^{2}}\bigl\{1-\text{erf}(d/\sqrt{2}%
\ell _{B})\bigr\}\biggr).  \label{BPpinStiff}
\end{equation}%
All isospins are spontaneously polarized to lower this exchange energy. As
far as the Hamiltonian (\ref{SU4Sigma}) concerns, there are six degenerate
states any one of which can be chosen as the ground state. It implies a
spontaneous breaking of the SU(4) symmetry, giving rise to Goldstone modes
and topological solitons.

The ground state is chosen actually by SU(4)-noninvariant interactions, as
described in the previous section. Goldstone modes are made gapful and
turned into pseudo-Goldstone modes. They are the soft waves in the system.
When the explicit breaking is sufficiently small, the pattern of the
spontaneous symmetry breaking provides us with an essential information on
soft waves.

We study a spontaneous symmetry breaking in the spin phase. The ground state
is a spin triplet and a ppin singlet. Let the spins be polarized to the $z$%
-direction. The residual symmetry is such one that keeps the ground state
invariant. One residual symmetry is the rotation about the spin-$z$ axis,
which is generated by the generator $S_{z}$. The ground state is also a ppin
SU(2) singlet. Thus, the rotation in the ppin space generated by $P_{a}$
keeps the ground state invariant. In addition to them the combined one $%
2S_{z}P_{a}$ does not transform the ground state. These seven
transformations generated by $S_{z},P_{a}$ and $2S_{z}P_{a}$ exhaust the
residual symmetry, which is $U(1)\otimes SU(2)\otimes SU(2)$. They form the
algebra,%
\begin{align}
& [S_{z},J_{a}]=0,~~~~~~[S_{z},K_{b}]=0,~~~~~[J_{a},J_{b}]=i\epsilon
_{abc}J_{c},  \notag \\
& [K_{a},K_{b}]=i\epsilon _{abc}K_{c},~~~~~~[J_{a},K_{b}]=0,
\end{align}%
where $J_{a}$ and $K_{a}$ are defined by 
\begin{equation}
J_{a}=\frac{1}{2}(P_{a}+2S_{z}P_{a}),\qquad K_{a}=\frac{1}{2}%
(P_{a}-2S_{z}P_{a}).
\end{equation}%
The pattern of the symmetry breaking is 
\begin{equation}
SU(4)\rightarrow U(1)\otimes SU(2)\otimes SU(2),  \label{SU4Break}
\end{equation}%
where the U(1) transformation is generated by $S_{z}$.

The target space is given by the coset space 
\begin{equation}
G_{4,2}=SU(4)/[U(1)\otimes SU(2)\otimes SU(2)].  \label{Grass42}
\end{equation}%
Here, $G_{N,k}$ is the complex Grassmannian manifold,%
\begin{equation}
G_{N,k}=U(N)/[U(k)\otimes U(N-k)].  \label{GrassNk}
\end{equation}%
The dimension of the manifold $G_{4,2}$ is $15-7=8$ corresponding to the
eight broken generators, $S_{x}$, $S_{y}$, $2S_{x}P_{a}$ and $2S_{y}P_{a}$.
Thus, there emerge eight Goldstone modes associated with them.

There is a nontrivial mapping to this coset space,

\begin{equation}
\pi _{2}\left( G_{N,k}\right) =\pi _{1}(U(1))=\mathbb{Z},  \label{SpinBreak}
\end{equation}%
where we have used $\pi _{2}(G/H)=\pi _{1}(H)$ (when G is simply connected)
and $\pi _{n}(G\otimes G^{\prime })=\pi _{n}(G)\oplus \pi _{n}(G^{\prime })$%
. Consequently, there emerge topological solitons indexed by the topological
number $\mathbb{Z}$. We call them $G_{4,2}$ skyrmions since they are
associated with the Grassmannian manifold $G_{4,2}$: See also section \ref%
{SecGrass}. Note that the $G_{4,1}$ skyrmion is the same object as the CP$%
^{3}$ skyrmion discussed extensively in the previous paper\cite{EzawaX02B}.
The nontrivial mapping characterizing the soliton is the U(1) group
generated by $S_{z}$.

Similarly we can study the spontaneous symmetry breaking in the ppin phase.
The ground state is a ppin triplet and a spin singlet. For simplicity we
consider the case, $d=0$, where all ppins are polarized to the $x$-direction
by the tunneling interaction. The pattern of the symmetry breaking is the
same as in the spin phase, as should be the case. The only difference is
that the nontrivial mapping is the U(1) group generated by $P_{x}$. The
seven residual symmetries are generated by $P_{x},S_{a}$ and $2S_{a}P_{x}$,
while the eight broken symmetries are generated by $P_{y}$, $P_{z}$, $%
2S_{a}P_{y}$ and $2S_{a}P_{z}$.

\section{Grassmannian fields}

\label{SecGrass}

There are fifteen generators in the nonlinear sigma model (\ref{SU4Sigma}),
but only eight of them are independent fields. To elucidate them we employ
the composite boson (CB) theory of QH ferromagnets\cite{Ezawa99L,Ezawa99JPSJ}
by attaching flux quanta to electrons\cite{Girvin87L,Read89L,Zhang89L}. The
CB field $\phi ^{\sigma }(\boldsymbol{x})$ is defined by making a singular phase
transformation to the electron field $\psi ^{\sigma }(\boldsymbol{x})$, 
\begin{equation}
\phi ^{\sigma }(\boldsymbol{x})=e^{-ie\Theta (\boldsymbol{x})}\psi ^{\sigma }(%
\boldsymbol{x}),  \label{NaiveCB}
\end{equation}%
where the phase field $\Theta (\boldsymbol{x})$ is subject to the relation%
\begin{equation}
\varepsilon _{ij}\partial _{i}\partial _{j}\Theta (\boldsymbol{x})=\Phi _{\text{D%
}}\rho (\boldsymbol{x}).  \label{PhaseCondi}
\end{equation}%
We introduce the normalized CB field $n^{\sigma }(\boldsymbol{x})$ by 
\begin{equation}
\phi ^{\sigma }(\boldsymbol{x})=\sqrt{\rho (\boldsymbol{x})}n^{\sigma }(\boldsymbol{x}),
\label{Skyrm2CB}
\end{equation}%
obeying%
\begin{equation}
\boldsymbol{n}^{\dagger }(\boldsymbol{x})\cdot \boldsymbol{n}(\boldsymbol{x})=\sum_{\sigma
}n^{\sigma \dagger }(\boldsymbol{x})n^{\sigma }(\boldsymbol{x})=1.  \label{BConstCp}
\end{equation}%
Such a field is the CP$^{3}$ field\cite{DAdda78NPB}.

At $\nu =2$ there are two electrons per one Landau site. Let us introduce
two CP$^{3}$ fields $\boldsymbol{n}_{1}(\boldsymbol{x})$ and $\boldsymbol{n}_{2}(\boldsymbol{%
x})$ for them. They should be orthogonal one to another, 
\begin{equation}
\boldsymbol{n}_{i}^{\dagger }(\boldsymbol{x})\cdot \boldsymbol{n}_{j}(\boldsymbol{x})=\delta
_{ij},
\end{equation}%
because they describe hard-core bosons. Using a set of two CP$^{3}$ fields
subject to this normalization condition we consider a $4\times 2$ matrix
field 
\begin{equation}
Z(\boldsymbol{x})=(\boldsymbol{n}_{1},\boldsymbol{n}_{2}),  \label{G42Field}
\end{equation}%
obeying 
\begin{equation}
Z^{\dagger }Z=1.
\end{equation}%
Though we have introduced two fields $\boldsymbol{n}_{1}(\boldsymbol{x})$ and $%
\boldsymbol{n}_{2}(\boldsymbol{x})$, we cannot distinguish them quantum mechanically
since they describe two electrons in the same Landau site. Namely, two
fields $Z(\boldsymbol{x})$ and $Z^{\prime }(\boldsymbol{x})$ are indistinguishable
physically when they are related by a local U(2) transformation $U(\boldsymbol{x}%
)$,%
\begin{equation}
Z^{\prime }(\boldsymbol{x})=Z(\boldsymbol{x})U(\boldsymbol{x}).
\end{equation}%
By identifying these two fields $Z(\boldsymbol{x})$ and $Z^{\prime }(\boldsymbol{x})$%
, the $4\times 2$ matrix field $Z(\boldsymbol{x})$ takes values on the
Grassmannian manifold $G_{4,2}$ defined by (\ref{Grass42}). The field $Z(%
\boldsymbol{x})$ is no longer a set of two independent CP$^{3}$ fields. It is a
new object, called the Grassmannian field, carrying eight real degrees of
freedom, as mentioned just below (\ref{GrassNk}).

\begin{figure}[tbph]
\includegraphics*[width=85mm]{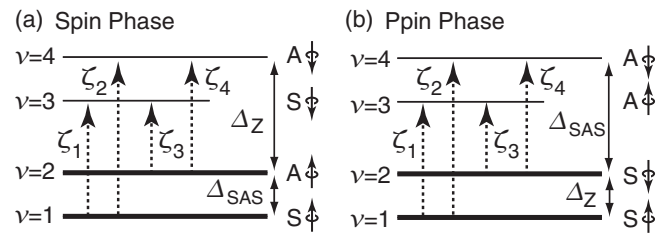}
\caption{The lowest two energy levels are occupied in the ground state of
the spin phase (a) and the ppin phase (b) at $\protect\nu=2$. Small
fluctuations are Goldstone modes $\protect\zeta_1$, $\protect\zeta_2$, $%
\protect\zeta_3$ and $\protect\zeta_4$. }
\label{BLvacc2PS}
\end{figure}

The lowest-energy one-body electron state is the up-spin symmetric state.
The second lowest energy state is either the up-spin antisymmetric state or
the down-spin symmetric state [Fig.\ref{BLvacc2PS}]. It is convenient to use
the CP$^{3}$ field whose components are taken in the symmetric-antisymmetric
basis, 
\begin{align}
n^{\text{S}\alpha }(\boldsymbol{x})& ={\frac{1}{\sqrt{2}}}\left( n^{\text{f}%
\alpha }(\boldsymbol{x})+n^{\text{b}\alpha }(\boldsymbol{x})\right) ,  \notag \\
n^{\text{A}\alpha }(\boldsymbol{x})& ={\frac{1}{\sqrt{2}}}\left( n^{\text{f}%
\alpha }(\boldsymbol{x})-n^{\text{b}\alpha }(\boldsymbol{x})\right) ,
\end{align}%
where $\alpha =\uparrow ,\downarrow $.

We first study the spin phase, where it is convenient to take the components
of the CP$^{3}$ field as [Fig.(a)]%
\begin{equation}
\boldsymbol{n}=(n^{\text{S}\uparrow },n^{\text{A}\uparrow },n^{\text{S}%
\downarrow },n^{\text{A}\downarrow }).
\end{equation}%
In the ground state all up-spin states are occupied, 
\begin{equation}
\boldsymbol{n}_{1}=(1,0,0,0),\quad \boldsymbol{n}_{2}=(0,1,0,0).  \label{SpinGroun}
\end{equation}%
Accordingly the ground state is represented by the $G_{4,2}$ field as 
\begin{equation}
Z_{\text{g}}=\left( 
\begin{array}{cc}
1 & 0 \\ 
0 & 1 \\ 
0 & 0 \\ 
0 & 0%
\end{array}%
\right) U(\boldsymbol{x}).  \label{G42Groun}
\end{equation}%
We study perturbations around this ground state. Up to the second order of
fluctuation fields, we may expand (\ref{G42Field}) as [Fig.\ref{BLvacc2PS}%
(a)] 
\begin{equation}
Z=\left( 
\begin{array}{cc}
1-\frac{1}{2}|\zeta _{1}|^{2}-\frac{1}{2}|\zeta _{2}|^{2} & -\frac{1}{2}%
\zeta _{1}^{\ast }\zeta _{3}-\frac{1}{2}\zeta _{2}^{\ast }\zeta _{4} \\ 
-\frac{1}{2}\zeta _{3}^{\ast }\zeta _{1}-\frac{1}{2}\zeta _{4}^{\ast }\zeta
_{2} & 1-\frac{1}{2}|\zeta _{3}|^{2}-\frac{1}{2}|\zeta _{4}|^{2} \\ 
\zeta _{1} & \zeta _{3} \\ 
\zeta _{2} & \zeta _{4}%
\end{array}%
\right) U(\boldsymbol{x}),  \label{G42Gold}
\end{equation}%
where $\zeta _{k}$ are the four complex Goldstone modes accompanied with the
spontaneous SU(4) breaking (\ref{SU4Break}). The Goldstone modes $\zeta _{1}$
and $\zeta _{4}$ are pure spin waves, while $\zeta _{2}$ and $\zeta _{3}$
are those mixing spins and pseudospins.

We can similarly analyze the ppin phase, where it is convenient to take the
components of the CP$^{3}$ field as [Fig.(b)]%
\begin{equation}
\boldsymbol{n}=(n^{\text{S}\uparrow },n^{\text{S}\downarrow },n^{\text{A}%
\uparrow },n^{\text{A}\downarrow })
\end{equation}%
In the ground state all up-spin states are occupied, 
\begin{equation}
\boldsymbol{n}_{1}=(1,0,0,0),\quad \boldsymbol{n}_{2}=(0,1,0,0).
\end{equation}%
The $G_{4,2}$ field is given by (\ref{G42Groun}), and the eight Goldstone
modes are parameterized as in (\ref{G42Gold}) with this choice of the
components of the CP$^{3}$ field.

\section{Soft Waves}

\label{SecSoftWave}

We have argued that the dynamic fields are given by the Grassmannian field $%
Z(\boldsymbol{x})$ in the BLQH system at $\nu =2$. On the other hand, in the
SU(4)-invariant limit the effective Hamiltonian is given by the nonlinear
SU(4) sigma model (\ref{SU4Sigma}) in terms of the SU(4) operator $\widehat{T%
}_{a}(\boldsymbol{x})$. By way of the relation 
\begin{equation}
\widehat{T}_{a}={\frac{1}{2}}\text{Tr}[Z^{\dagger }\lambda _{a}Z]={\frac{1}{2%
}}\boldsymbol{n}_{1}^{\dagger }\lambda _{a}\boldsymbol{n}_{1}+{\frac{1}{2}}\boldsymbol{n}%
_{2}^{\dagger }\lambda _{a}\boldsymbol{n}_{2},  \label{SpinToZ}
\end{equation}%
we are able to rewrite it into the well-known Hamiltonian\cite%
{MacFarlane82PLB} for the Grassmannian field, 
\begin{equation}
\mathcal{H}_{X}^{+}=2J^{+}\text{Tr}\left[ (\partial _{j}Z-iZK_{j})^{\dagger
})(\partial _{j}Z-iZK_{j})\right] ,  \label{ExchaGrass}
\end{equation}%
where 
\begin{equation}
K_{\mu }(\boldsymbol{x})=-iZ^{\dagger }(\boldsymbol{x})\partial _{\mu }Z(\boldsymbol{x}).
\label{AuxiaK}
\end{equation}%
This Hamiltonian has the local U(2) gauge symmetry, 
\begin{subequations}
\begin{align}
Z(\boldsymbol{x})& \rightarrow Z(\boldsymbol{x})U(\boldsymbol{x}), \\
K_{\mu }(\boldsymbol{x})& \rightarrow U(\boldsymbol{x})^{\dagger }K_{\mu }(\boldsymbol{x}%
)U(\boldsymbol{x})-iU(\boldsymbol{x})^{\dagger }\partial _{\mu }U(\boldsymbol{x}).
\label{LocalU2}
\end{align}%
The gauge field $K_{\mu }$ is not a dynamic field since it is an auxiliary
field given by (\ref{AuxiaK}).

We study small fluctuations of the soft waves in the spin phase.
Substituting the parametrization (\ref{G42Gold}) of the Grassmannian field
into the Hamiltonian (\ref{ExchaGrass}), we expand it up to the second
order, 
\end{subequations}
\begin{eqnarray}
\mathcal{H}_{X}^{+} &=&{2J^{+}}\sum_{i=1}^{4}\partial _{k}\zeta
_{i}^{\dagger }(\boldsymbol{x})\partial _{k}\zeta _{i}(\boldsymbol{x})  \notag \\
&=&{\frac{J^{+}}{2}}\sum_{i=1}^{4}\bigl\{(\partial _{k}\sigma
_{i})^{2}+(\partial _{k}\vartheta _{i})^{2}\bigr\},  \label{EffecHamilSpin}
\end{eqnarray}%
where 
\begin{equation}
\zeta _{i}(\boldsymbol{x})={\frac{1}{2}}\bigl(\sigma _{i}(\boldsymbol{x})+i\vartheta
_{i}(\boldsymbol{x})\bigr).  \label{ZetaField}
\end{equation}%
The equal-time commutation relation is%
\begin{equation}
\lbrack \sigma _{i}(\boldsymbol{x}),\vartheta _{j}(\boldsymbol{y})]=2i\rho _{\Phi
}^{-1}\delta _{ij}\delta (\boldsymbol{x}-\boldsymbol{y}),
\end{equation}%
where $\rho _{\Phi }=\frac{1}{2}\rho _{0}=1/(2\pi \ell _{B}^{2})$ represents
the flux density. This Hamiltonian realizes the SU(4) symmetry nonlinearly.
It describes four Goldstone modes associated with spontaneous symmetry
breakdown of the SU(4) symmetry.

The SU(4) symmetry is broken explicitly but softly by various direct
interactions. Relevant SU(2) operators are 
\begin{subequations}
\begin{align}
\widehat{S}_{z}& ={\frac{1}{2}}\text{Tr}\left[ Z^{\dagger }\left( 
\begin{array}{cc}
1_{2} & 0 \\ 
0 & -1_{2}%
\end{array}%
\right) Z\right] \simeq 1-\sum_{j=1}^{4}|\zeta _{j}|^{2}, \\
\widehat{P}_{x}& ={\frac{1}{2}}\text{Tr}\left[ Z^{\dagger }\left( 
\begin{array}{cc}
1_{2} & 0 \\ 
0 & 1_{2}%
\end{array}%
\right) Z\right] \simeq -|\zeta _{1}|^{2}+|\zeta _{3}|^{2},  \label{SPP} \\
\widehat{P}_{z}& ={\frac{1}{2}}\text{Tr}\left[ Z^{\dagger }\left( 
\begin{array}{cc}
\tau _{1} & 0 \\ 
0 & \tau _{1}%
\end{array}%
\right) Z\right] \simeq \text{Re}[(\zeta _{1}-\zeta _{4})(\zeta _{2}^{\ast
}-\zeta _{3}^{\ast })],
\end{align}%
up to the second order of fluctuation fields, where the $4\times 4$ matrices
are taken in the symmetric-asymmetric basis in accord with (\ref{G42Gold}).

By taking into account of the SU(4)-noninvariant exchange interaction as
well, the effective Hamiltonian is decomposed into four independent modes, $%
\mathcal{H}_{X}=\sum_{i=1}^{4}\mathcal{H}_{i}$, where 
\end{subequations}
\begin{subequations}
\begin{equation}
\mathcal{H}_{1}={\frac{J}{2}}\bigl\{(\partial _{k}\sigma _{1})^{2}+(\partial
_{k}\vartheta _{1})^{2}\bigr\}+{\frac{{\Delta }_{\text{Z}}\rho _{\Phi }}{4}}%
\bigl(\sigma _{1}^{2}+\vartheta _{1}^{2}\bigr),  \label{SpinWave1}
\end{equation}%
\begin{equation}
\mathcal{H}_{4}={\frac{J}{2}}\bigl\{(\partial _{k}\sigma _{4})^{2}+(\partial
_{k}\vartheta _{4})^{2}\bigr\}+{\frac{{\Delta }_{\text{Z}}\rho _{\Phi }}{4}}%
\bigl(\sigma _{4}^{2}+\vartheta _{4}^{2}\bigr),  \label{SpinWave2}
\end{equation}%
\begin{equation}
\mathcal{H}_{2}={\frac{J^{d}}{2}}\bigl\{(\partial _{k}\sigma
_{2})^{2}+(\partial _{k}\vartheta _{2})^{2}\bigr\}+{\frac{({\Delta }_{\text{Z%
}}+{\Delta }_{\text{SAS}})\rho _{\Phi }}{4}}\bigl(\sigma _{2}^{2}+\vartheta
_{2}^{2}\bigr),  \label{IpinWave1}
\end{equation}%
\begin{equation}
\mathcal{H}_{3}={\frac{J^{d}}{2}}\bigl\{(\partial _{k}\sigma
_{3})^{2}+(\partial _{k}\vartheta _{3})^{2}\bigr\}+{\frac{|{\Delta }_{\text{Z%
}}-{\Delta }_{\text{SAS}}|\rho _{\Phi }}{4}}\bigl(\sigma _{3}^{2}+\vartheta
_{3}^{2}\bigr).  \label{IpinWave2}
\end{equation}%
They describe four independent soft waves, which are pseudo-Goldstone modes
by acquiring gaps. They have finite coherence lengths (correlation lengths), 
\end{subequations}
\begin{subequations}
\begin{align}
\xi _{1}& =\xi _{4}=\ell _{B}\sqrt{{\frac{4\pi J}{{\Delta }_{\text{Z}}}}},
\label{SpinLengt} \\
\xi _{\text{2}}& =\ell _{B}\sqrt{{\frac{4\pi J^{d}}{{\Delta }_{\text{Z}}+{%
\Delta }_{\text{SAS}}}}},  \label{IpinLengt1} \\
\xi _{\text{3}}& =\ell _{B}\sqrt{{\frac{4\pi J^{d}}{|{\Delta }_{\text{Z}}-{%
\Delta }_{\text{SAS}}|}}}.  \label{IpinLengt2}
\end{align}%
The two modes $\zeta _{1}$ and $\zeta _{4}$ are degenerate. The mode $\zeta
_{\text{3}}$ becomes gapless at ${\Delta }_{\text{Z}}={\Delta }_{\text{SAS}}$%
, which signals the breakdown of the spin phase due to an infrared
catastrophe. As $d\rightarrow \infty $ we find $\xi _{2}\rightarrow 0$ and $%
\xi _{3}\rightarrow 0$ since $J^{d}\rightarrow 0$, which implies the
disappearance of the Goldstone modes $\zeta _{2}$ and $\zeta _{3}$ due to
the decoupling of the bilayer system into the two monolayer systems.

We may perform a similar analysis in the ppin phase. Relevant operators for
the direct interactions are 
\end{subequations}
\begin{subequations}
\begin{align}
\hat{S}_{z}& \simeq -|\zeta _{2}|^{2}+|\zeta _{3}|^{2}, \\
\hat{P}_{x}& \simeq 1-(|\zeta _{1}|^{2}+|\zeta _{2}|^{2}+|\zeta
_{3}|^{2}+|\zeta _{4}|^{2}), \\
\hat{P}_{z}& \simeq \frac{1}{2}(\zeta _{1}^{\dagger }+\zeta _{1}+\zeta
_{4}^{\dagger }+\zeta _{4}).
\end{align}%
The effective Hamiltonian is decomposed into four independent modes, $%
\mathcal{H}_{\text{X}}=\sum_{i=1}^{4}\mathcal{H}_{i}$. The Hamiltonians $%
\mathcal{H}_{2}$ and $\mathcal{H}_{3}$ are given by the same Hamiltonians (%
\ref{IpinWave1}) and (\ref{IpinWave2}), respectively. The coherence lengths
are given by (\ref{IpinLengt1}) and (\ref{IpinLengt2}). The mode $\zeta _{%
\text{3}}$ becomes gapless at ${\Delta }_{\text{Z}}={\Delta }_{\text{SAS}}$,
which signals the breakdown of the ppin phase due to an infrared
catastrophe. On the other hand, the Hamiltonian $\mathcal{H}_{1}$ reads 
\end{subequations}
\begin{align}
\mathcal{H}_{1}=& {\frac{J}{2}}(\partial _{k}\sigma _{1})^{2}+{\frac{J^{d}}{2%
}}(\partial _{k}\vartheta _{1})^{2}+{\frac{\varepsilon _{\text{cap}}\rho
_{\Phi }}{4}}\sigma _{1}^{2}  \notag \\
& +{\frac{{\Delta }_{\text{SAS}}\rho _{\Phi }}{4}}\bigl(\sigma
_{1}^{2}+\vartheta _{1}^{2}\bigr)\boldsymbol{\ }.  \label{PpinWave}
\end{align}%
The Hamiltonians $\mathcal{H}_{4}$ is given by the same Hamiltonian as $%
\mathcal{H}_{1}$ with the replacement of the fields $(\sigma _{1},\vartheta
_{1})$ by $(\sigma _{4},\vartheta _{4})$. There are two coherent lengths
associated with the $\sigma $ field and the $\vartheta $ field, 
\begin{subequations}
\begin{align}
\xi _{1}^{\vartheta }& =\xi _{4}^{\vartheta }=\ell _{B}\sqrt{{\frac{4\pi
J^{d}}{{\Delta }_{\text{SAS}}}}}, \\
\xi _{1}^{\sigma }& =\xi _{4}^{\sigma }=\ell _{B}\sqrt{{\frac{4\pi J}{%
\varepsilon _{\text{cap}}+{\Delta }_{\text{SAS}}}}}.
\end{align}%
The ground state of these modes (\ref{PpinWave}) is a squeezed state as in
the $\nu =1$ case\cite{EzawaX02B}, where the coherence lengths are different
between the conjugate variables. A new feature is that the interlayer
tunneling modes $\xi _{1}$ and $\xi _{4}$ mediate the Josephson-like effect
as in the $\nu =1$ case\cite{Ezawa92IJMPB,BookEzawa}.

\section{Grassmannian $G_{4,2}$ Skyrmions}

\label{SecGrassSolit}

The homotopy theorem (\ref{SpinBreak}) guarantees the existence of
topological solitons ($G_{4,2}$ skyrmions). The topological charge is defined%
\cite{MacFarlane82PLB} as a gauge invariant by 
\end{subequations}
\begin{equation}
Q={\frac{i}{2\pi }}\int d^{2}x\epsilon _{jk}\text{Tr}\left[ (\partial
_{j}Z-iZK_{j})^{\dagger }(\partial _{k}Z-iZK_{k})\right] .
\label{TopolCharg}
\end{equation}%
It is a topological invariant since it is equal to 
\begin{equation}
Q={\frac{i}{2\pi }}\int d^{2}x\epsilon _{jk}\text{Tr}\left[ (\partial
_{j}Z)^{\dagger }(\partial _{k}Z)\right] .
\end{equation}%
We rewrite it with the use of (\ref{G42Field}) as%
\begin{equation}
Q={\frac{i}{2\pi }}\int d^{2}x\epsilon _{jk}\left[ (\partial _{j}\boldsymbol{n}%
_{1})^{\dagger }\cdot (\partial _{k}\boldsymbol{n}_{1})+(\partial _{j}\boldsymbol{n}%
_{2})^{\dagger }\cdot (\partial _{k}\boldsymbol{n}_{2})\right] .
\label{TopolCharg2CP}
\end{equation}%
It is the sum of the topological charges associated with the CP$^{3}$ fields 
$\boldsymbol{n}_{1}$ and $\boldsymbol{n}_{2}.$Hence, the $G_{4,2}$ skyrmion consists
of CP$^{3}$ skyrmions,%
\begin{equation}
n_{k}^{\sigma }(\boldsymbol{x})=\frac{1}{\sqrt{\sum |\omega _{k}^{\sigma }(z)|}}%
\omega _{k}^{\sigma }(z),  \label{EachCP}
\end{equation}%
excited in the front and back layers in the spin phase, or in the up-spin
and down-spin states in the ppin phase. Here, $\omega _{k}^{\sigma }(z)$ are
arbitrary analytic functions.

For definiteness we consider the spin phase, where we now choose $\boldsymbol{n}%
_{k}=(n_{k}^{\text{f}\uparrow },n_{k}^{\text{b}\uparrow },n_{k}^{\text{f}%
\downarrow },n_{k}^{\text{b}\downarrow })$. The simplest soliton would be a
set of one CP$^{3}$ skyrmion in the front layer and the ground state in the
back layer, $\boldsymbol{n}_{1}=(z,\kappa ,0,0)/\sqrt{|z|^{2}+\kappa ^{2}}$ and $%
\boldsymbol{n}_{2}=(0,1,0,0)$, or%
\begin{equation}
Z_{\text{sky}}^{1}=\frac{1}{\sqrt{|z|^{2}+\kappa ^{2}}}\left( 
\begin{array}{cc}
z & 0 \\ 
0 & \sqrt{|z|^{2}+\kappa ^{2}} \\ 
\kappa & 0 \\ 
0 & 0%
\end{array}%
\right) U(\boldsymbol{x}),  \label{G42Sky1}
\end{equation}%
for which the topological charge (\ref{TopolCharg}) is $Q=1$. The next
simplest would be a set of two CP$^{3}$ skyrmions both in the front and back
layers,%
\begin{equation}
Z_{\text{sky}}^{2}=\frac{1}{\sqrt{|z|^{2}+\kappa ^{2}}}\left( 
\begin{array}{cc}
z & 0 \\ 
0 & z \\ 
\kappa & 0 \\ 
0 & k%
\end{array}%
\right) U(\boldsymbol{x}),  \label{G42Sky2}
\end{equation}%
for which $Q=2$. We now argue that the simplest G$_{4,2}$ skyrmion (\ref%
{G42Sky1}) is ruled out since it is not confined within the lowest Landau
level.

The Landau-site Hamiltonian (\ref{SU4Site}) has been derived by requiring
the LLL condition. However, the nonlinear sigma model (\ref{SU4Sigma}),
obtained after taking a continuum limit, is an ordinary local Hamiltonian,
and the CP$^{N-1}$ field (\ref{Skyrm2CB}) is an ordinary field without the
LLL projection. Hence, it is necessary to require the LLL condition on the
soliton states\cite{BookEzawa}.

The condition requires the kinetic Hamiltonian to be quenched on the state,
which reads\cite{Ezawa99L}%
\begin{equation}
\frac{\partial }{\partial z^{\ast }}\varphi ^{\sigma }(\boldsymbol{x})|\Phi
\rangle =0.  \label{CondiLLL}
\end{equation}%
Here $\varphi ^{\sigma }(\boldsymbol{x})$ is the dressed CB field\cite%
{Ezawa99L,Read89L} defined by%
\begin{equation}
\varphi ^{\sigma }(\boldsymbol{x})=e^{-\mathcal{A}(\boldsymbol{x})}\phi ^{\sigma }(%
\boldsymbol{x}),  \label{DressCB}
\end{equation}%
where $\phi ^{\sigma }(\boldsymbol{x})$ is the CB field (\ref{NaiveCB}), while ${%
\mathcal{A}}(\boldsymbol{x})$ is an auxiliary field determined by%
\begin{equation}
\boldsymbol{\nabla }^{2}{\mathcal{A}}(\boldsymbol{x})=2\pi \left( \rho (\boldsymbol{x}%
)-\rho _{0}\right) ,  \label{Dress2OperaE}
\end{equation}%
as follows from the condition (\ref{PhaseCondi}) on the phase field. We take
a coherent state of $\varphi ^{\sigma }(\boldsymbol{x})$, for which (\ref%
{CondiLLL}) implies%
\begin{equation*}
\varphi ^{\sigma }(\boldsymbol{x})|\Phi \rangle =\omega ^{\sigma }(z)|\Phi
\rangle ,
\end{equation*}%
where $\omega ^{\sigma }(z)$ is an analytic function. The coherent state $%
|\Phi \rangle $ must be an eigenstate of the density operator $\rho (\boldsymbol{%
x})$ and a coherent state of the CP$^{3}$ field $\boldsymbol{n}(\boldsymbol{x})$
since they commute with each other. Hence we have 
\begin{equation}
e^{-\mathcal{A}^{\text{cl}}(\boldsymbol{x})}\sqrt{\rho ^{\text{cl}}(\boldsymbol{x})}%
n^{\text{cl(}\sigma )}(\boldsymbol{x})=\omega ^{\sigma }(z),  \label{Dress2CB}
\end{equation}%
where $\mathcal{A}^{\text{cl}}(\boldsymbol{x})$, $\rho ^{\text{cl}}(\boldsymbol{x})$
and $n^{\text{cl(}\sigma )}(\boldsymbol{x})$ are classical fields. This is the
LLL condition for soliton states.

At $\nu =2$ we have 
\begin{equation}
\boldsymbol{n}(\boldsymbol{x})|\Phi \rangle =\boldsymbol{n}^{\text{cl}}(\boldsymbol{x})|\Phi
\rangle =\left[ \boldsymbol{n}_{1}^{\text{cl}}(\boldsymbol{x})+\boldsymbol{n}_{2}^{\text{%
cl}}(\boldsymbol{x})\right] |\Phi \rangle ,  \label{TwoCP}
\end{equation}%
together with $\boldsymbol{n}_{i}^{\text{cl}}(\boldsymbol{x})\cdot \boldsymbol{n}_{j}^{%
\text{cl}}(\boldsymbol{x})=\delta _{ij}$ and 
\begin{equation}
\rho (\boldsymbol{x})|\Phi \rangle =\rho ^{\text{cl}}(\boldsymbol{x})|\Phi \rangle
=[\rho _{1}^{\text{cl}}(\boldsymbol{x})+\rho _{2}^{\text{cl}}(\boldsymbol{x})]|\Phi
\rangle .
\end{equation}%
Keeping in mind the local U(2) invariance we work in such a gauge that $U(%
\boldsymbol{x})=1$ in (\ref{G42Sky1}) and (\ref{G42Sky2}). We may solve (\ref%
{Dress2CB}) as 
\begin{equation}
n^{\text{cl(}\sigma )}(\boldsymbol{x})=\frac{\sqrt{2}}{\sqrt{\sum |\omega
^{\sigma }(x)|^{2}}}\omega ^{\sigma }(z).  \label{CondiCP}
\end{equation}%
Comparing this with (\ref{EachCP}) we conclude $\sum_{\sigma }|\omega
_{1}^{\sigma }(z)|^{2}=\sum_{\sigma }|\omega _{2}^{\sigma }(z)|^{2}$ and $%
\omega ^{\sigma }(z)=\omega _{1}^{\sigma }(z)+\omega _{2}^{\sigma }(z)$.
Therefore, for each component the LLL condition (\ref{Dress2CB}) holds and
we obtain from (\ref{Dress2OperaE}) that 
\begin{equation}
{\frac{1}{4\pi }}\boldsymbol{\nabla }^{2}\ln \rho ^{\text{cl}}(\boldsymbol{x})-\rho
^{\text{cl}}(\boldsymbol{x})+\rho _{0}=j_{\text{sky}}^{0}(\boldsymbol{x})
\label{SolitEq}
\end{equation}%
with 
\begin{eqnarray}
j_{\text{sky}}^{0}(\boldsymbol{x}) &=&{\frac{1}{4\pi }}\boldsymbol{\nabla }^{2}\ln
\sum_{\sigma }|\omega _{1}^{\sigma }(z)|^{2}  \notag \\
&=&{\frac{1}{4\pi }}\boldsymbol{\nabla }^{2}\ln \sum_{\sigma }|\omega
_{2}^{\sigma }(z)|^{2}.  \label{Semi2ClassCharg}
\end{eqnarray}%
It is easy to see that the topological charge (\ref{TopolCharg2CP}) is given
by $Q=2\int d^{2}x\,j_{\text{sky}}^{0}(\boldsymbol{x})$.

The G$_{4,2}$ skyrmion has a general expression,%
\begin{equation}
Z_{\text{sky}}=\frac{1}{\sqrt{\sum_{\sigma }|\omega _{1}^{\sigma }(z)|^{2}}}%
\left( 
\begin{array}{cc}
\omega _{1}^{\text{f}\uparrow }(z) & \omega _{2}^{\text{f}\uparrow }(z) \\ 
\omega _{1}^{\text{f}\downarrow }(z) & \omega _{2}^{\text{f}\downarrow }(z)
\\ 
\omega _{1}^{\text{b}\uparrow }(z) & \omega _{2}^{\text{b}\uparrow }(z) \\ 
\omega _{1}^{\text{b}\downarrow }(z) & \omega _{2}^{\text{b}\downarrow }(z)%
\end{array}%
\right) .  \label{GenerBisky}
\end{equation}%
This rules out the soliton (\ref{G42Sky1}) with $Q=1$. The simplest G$_{4,2}$
skyrmion is given by (\ref{G42Sky2}), which describes a pair of CP$^{3}$
skyrmions excited in both layers. We call it a biskyrmion.

We emphasize this peculiar situation by recalling the formula (\ref{Skyrm2CB}%
) to introduce the normalized CB field $\boldsymbol{n(x)}$. It is essential that
the total electron density $\rho (\boldsymbol{x})$ is common to all the four
components: Otherwise the SU(4) symmetry is explicitly broken by hand. Even
if we try to excite a skyrmion only in the front layer, the density
modulation associated with it affects equally electrons in the back layer as
far as the LLL condition (\ref{Dress2CB}) is respected. In a genuine BLQH
system it is impossible to have a skyrmion excitation in the front layer and
the ground state in the back layer simultaneously: See section \ref%
{SecTwoMono}\ for more details on this point.

Finally we note that the G$_{4,2}$ skyrmion is a BPS soliton of the exchange
Hamiltonian (\ref{ExchaGrass}). Indeed, the following inequality\cite%
{MacFarlane82PLB} holds between the exchange energy (\ref{ExchaGrass}) and
the topological charge (\ref{TopolCharg}),%
\begin{equation}
E_{X}\geqq 4\pi J^{+}Q,  \label{CondiBPS}
\end{equation}%
where the equality is achieved by the G$_{4,2}$ skyrmion (\ref{GenerBisky}).

A completely analogous analysis is made in the ppin phase by choosing $%
\boldsymbol{n}=(n^{\text{S}\uparrow },n^{\text{S}\downarrow },n^{\text{A}%
\uparrow },n^{\text{A}\downarrow })$. We achieve the same conclusion that
the lightest topological excitation is a biskyrmion.

\section{Spin-1 Sigma Models for Biskyrmions}

\label{SecEffecTheor}

We have argued that charged excitations are biskyrmions in the $\nu =2$ BLQH
system. We have described them in terms of the Grassmannian field $Z(\boldsymbol{%
x}).$ We now represent them in terms of the SU(4) sigma field by way of the
relation (\ref{SpinToZ}).

We first treat the spin phase. We calculate the SU(4) generators $\widehat{T}%
_{a}(\boldsymbol{x})$ by using the biskyrmion configuration (\ref{G42Sky2}) in (%
\ref{SpinToZ}), from which the fifteen operators $\widehat{S}_{a}$, $%
\widehat{P}_{a}$ and $2\widehat{S}_{a}\widehat{P}_{b}$ are derived with the
aid of the relations (\ref{T2SP}). We find that $\widehat{P}_{a}=2\widehat{S}%
_{a}\widehat{P}_{b}=0$ and that $\widehat{S}_{a}$ are given by the
well-known formula of the O(3) skyrmion,%
\begin{align}
\widehat{S}_{x}& =\frac{2\kappa |z|}{|z|^{2}+\kappa ^{2}}\cos \theta ,\quad 
\widehat{S}_{y}=-\frac{2\kappa |z|}{|z|^{2}+\kappa ^{2}}\sin \theta ,  \notag
\\
\widehat{S}_{z}& =\frac{|z|^{2}-\kappa ^{2}}{|z|^{2}+\kappa ^{2}}.
\label{SpinSkyrm}
\end{align}%
It is easy to see that this biskyrmion configuration is purely spin-like and
expanded by three states (\ref{GrounT}). We call it the spin biskyrmion.

We define bosonic operators $t_{a}$ on the lattice by 
\begin{equation}
t_{a}^{\dagger }(i)|0\rangle =|t_{a}\rangle _{i}.  \label{BosonT}
\end{equation}%
They satisfy hard-core bosonic commutation relations and describe Schwinger
bosons\cite{Sachdev90B}. In the field-theoretical limit we have%
\begin{align}
t_{\uparrow }(\boldsymbol{x})& =n^{\text{f}\uparrow }(\boldsymbol{x})n^{\text{b}%
\uparrow }(\boldsymbol{x}),  \notag \\
t_{0}(\boldsymbol{x})& =\frac{1}{\sqrt{2}}\bigl(n^{\text{f}\uparrow }(\boldsymbol{x}%
)n^{\text{b}\downarrow }(\boldsymbol{x})+n^{\text{f}\downarrow }(\boldsymbol{x})n^{%
\text{b}\uparrow }(\boldsymbol{x})\bigr),  \notag \\
t_{\downarrow }(\boldsymbol{x})& =n^{\text{f}\downarrow }(\boldsymbol{x})n^{\text{b}%
\downarrow }(\boldsymbol{x}),  \label{Spin1S}
\end{align}%
which read%
\begin{equation}
\boldsymbol{t}(\boldsymbol{x})=(t_{\uparrow },t_{0},t_{\downarrow })=\frac{1}{%
|z|^{2}+\kappa ^{2}}(z^{2},\sqrt{2}z\kappa ,\kappa ^{2})  \label{SkyrmCP}
\end{equation}%
on the biskyrmion configuration (\ref{G42Sky2}). The spin-1 field is given by

\begin{equation}
\widehat{S}_{a}(\boldsymbol{x})=\boldsymbol{t}^{\dagger }(\boldsymbol{x})\widehat{\tau }%
_{a}\boldsymbol{t}(\boldsymbol{x}),  \label{CPspin}
\end{equation}%
where $\widehat{\tau }_{a}$ are the SU(2) generators in the adjoint
representation, 
\begin{equation}
\widehat{\boldsymbol{\tau }}\equiv \biggl\{\frac{1}{\sqrt{2}}%
\begin{pmatrix}
0 & 1 & 0 \\ 
1 & 0 & 1 \\ 
0 & 1 & 0%
\end{pmatrix}%
,\frac{1}{\sqrt{2}}%
\begin{pmatrix}
0 & {-i} & 0 \\ 
{i} & 0 & {-i} \\ 
0 & {i} & 0%
\end{pmatrix}%
,%
\begin{pmatrix}
1 & 0 & 0 \\ 
0 & 0 & 0 \\ 
0 & 0 & -1%
\end{pmatrix}%
\biggr\}.  \label{ExactParam}
\end{equation}%
Calculating (\ref{CPspin}) with (\ref{Spin1S}) and (\ref{SkyrmCP}), we
reproduce the skyrmion configuration (\ref{SpinSkyrm}). Hence, the
Grassmannian G$_{4,2}$ skyrmion (\ref{G42Sky2}) in the spin phase is equal
to the O(3) skyrmion in the spin sector.

We reduce the Hamiltonian (\ref{Ex}) to the spin sector. Because $\langle 
\widehat{P}_{a}\rangle =\langle \widehat{S}_{a}\widehat{P}_{b}\rangle =0$,
we obtain that%
\begin{equation}
H_{X}=-2\sum_{\langle i,j\rangle }J_{ij}\widehat{\boldsymbol{S}}(i)\!\cdot \!%
\widehat{\boldsymbol{S}}(j).
\end{equation}%
Taking the continuum limit and including the direct interactions, we find%
\begin{equation}
{\mathcal{H}}_{\text{spin}}=J\partial _{k}\widehat{\boldsymbol{S}}(\boldsymbol{x}%
)\!\cdot \!\partial _{k}\widehat{\boldsymbol{S}}(\boldsymbol{x})+{\mathcal{H}}_{%
\text{Coulomb}}+{\Delta }_{\text{Z}}\bar{\rho}_{B}\left( 1-\widehat{S}_{z}(%
\boldsymbol{x})\right) ,  \label{ContiSpin}
\end{equation}%
where ${\mathcal{H}}_{\text{Coulomb}}$ stands for the direct Coulomb energy.
The O(3) skyrmion (\ref{SpinSkyrm}) is the BPS soliton of the
O(3)-nonlinear-sigma-model part of this Hamiltonian. The total Hamiltonian (%
\ref{ContiSpin}) is very similar to the one in the monolayer QH system at $%
\nu =1$ with the replacement of the spin-${\frac{1}{2}}$ field by the spin-$%
1 $ field.

The excitation energy of the biskyrmion is easily calculable based on (\ref%
{ContiSpin}) as in the $\nu =1$ monolayer QH system \cite{Ezawa99JPSJ}. The
skyrmion scale $\kappa $ is determined to optimize the Coulomb energy and
the Zeeman energy. It is to be remarked that the spin biskyrmion is
insensible to the interlayer stiffness $J^{d}$.

We may similarly discuss the ppin phase, where only $\widehat{P}_{a}$ is
nonvanishing for the biskyrmion configuration. The biskyrmion turns out to
be the O(3) skyrmion,%
\begin{align}
\widehat{P}_{y}& =\frac{2\kappa |z|}{|z|^{2}+\kappa ^{2}}\cos \theta ,\quad 
\widehat{P}_{z}=-\frac{2\kappa |z|}{|z|^{2}+\kappa ^{2}}\sin \theta ,  \notag
\\
\widehat{P}_{x}& =\frac{|z|^{2}-\kappa ^{2}}{|z|^{2}+\kappa ^{2}}.
\label{PpinSkyrm}
\end{align}%
We call it the ppin biskyrmion. The effective Hamiltonian restricted to the
ppin sector is%
\begin{align}
{\mathcal{H}}_{\text{ppin}}& =J^{d}\bigl\{\partial _{k}\widehat{\boldsymbol{P}}(%
\boldsymbol{x})\!\cdot \!\partial _{k}\widehat{\boldsymbol{P}}(\boldsymbol{x})\bigr\}%
_{xy}+J\partial _{k}\widehat{P}_{z}(\boldsymbol{x})\cdot \partial _{k}\widehat{P}%
_{z}(\boldsymbol{x})  \notag \\
& +{\mathcal{H}}_{\text{Coulomb}}+{\Delta }_{\text{SAS}}\bar{\rho}_{B}\left(
1-\widehat{P}_{x}(\boldsymbol{x})\right) ,  \label{ContiPpin}
\end{align}%
where ${\mathcal{H}}_{\text{Coulomb}}$ stands for the direct Coulomb energy
including the capacitance effect. The total Hamiltonian is quite similar to
that in the spin-frozen BLQH system at $\nu =1$ with the replacement of the
ppin-${\frac{1}{2}}$ field with the ppin-$1$ field.

\section{Two Monolayer Systems}

\label{SecTwoMono}

It is intriguing that a quasiparticle is a biskyrmion at $\nu =2$. However,
it is clear intuitively that a quasiparticle is a simple skyrmion excited in
one of the two layers even at $\nu =2$ if the two layers are sufficiently
separated. We have studied previously\cite{EzawaX02B,RJ} the criterion for a
system at $\nu =1$ to be a genuine bilayer system or a set of two monolayer
systems. It is determined by the local symmetry present in the Hamiltonian.
Let us recapitulate the argument\cite{EzawaX02B} and extend it to the system
at $\nu =2$.

We first examine the local symmetry of the direct interaction (\ref{BDct}).
It is given by a direct product of two U(1) symmetries, U$^{\uparrow }$(1)$%
\otimes $U$^{\downarrow }$(1), 
\begin{align}
\begin{pmatrix}
\psi ^{\text{f}\uparrow }(\boldsymbol{x}) \\ 
\psi ^{\text{b}\uparrow }(\boldsymbol{x})%
\end{pmatrix}%
& \longrightarrow e^{i\alpha (\boldsymbol{x})}%
\begin{pmatrix}
\psi ^{\text{f}\uparrow }(\boldsymbol{x}) \\ 
\psi ^{\text{b}\uparrow }(\boldsymbol{x})%
\end{pmatrix}%
,  \notag \\
\begin{pmatrix}
\psi ^{\text{f}\downarrow }(\boldsymbol{x}) \\ 
\psi ^{\text{b}\downarrow }(\boldsymbol{x})%
\end{pmatrix}%
& \longrightarrow e^{i\beta (\boldsymbol{x})}%
\begin{pmatrix}
\psi ^{\text{f}\downarrow }(\boldsymbol{x}) \\ 
\psi ^{\text{b}\downarrow }(\boldsymbol{x})%
\end{pmatrix}%
.
\end{align}%
The exchange interaction (\ref{Ex}) breaks this into a single U(1) symmetry, 
\begin{equation}
\psi ^{\sigma }(\boldsymbol{x})\longrightarrow e^{i\alpha (\boldsymbol{x})}\psi
^{\sigma }(\boldsymbol{x}).  \label{B2LocalU1}
\end{equation}%
Note that this is the case even if $J^{d}=0$ provided $J\neq 0$. It is the
exact local symmetry of the total Hamiltonian. Corresponding to this U(1)
symmetry, we have introduced the normalized CB field $n^{\sigma }(\boldsymbol{x}%
) $ in (\ref{Skyrm2CB}), or 
\begin{equation}
\phi ^{\sigma }(\boldsymbol{x})=\sqrt{\rho (\boldsymbol{x})}n^{\sigma }(\boldsymbol{x}).
\label{OneCP3}
\end{equation}%
It is the CP$^{3}$ field containing 3 independent complex fields, because
one real field is eliminated by the constraint (\ref{BConstCp}) and
furthermore the U(1) phase field is not dynamic due to the local U(1)
symmetry (\ref{B2LocalU1}). At $\nu =2$ we introduce a set of two CP$^{3}$
fields for two electrons in one Landau site with the U(2) local symmetry (%
\ref{LocalU2}). The set turns out to be a Grassmannian G$_{4,2}$ field with
four independent complex fields. Topological solitons are biskyrmions.

We next consider a system where the two layers are separated sufficiently so
that there are no interlayer exchange interaction ($J^{d}=0$) nor the
tunneling interaction (${\Delta }_{\text{SAS}}=0$). Then, the total
Hamiltonian is invariant under two local transformations, U$^{\text{f}}$(1)
and U$^{\text{b}}$(1), which act on electrons on the two layers
independently, 
\begin{align}
\begin{pmatrix}
\psi ^{\text{f}\uparrow }(\boldsymbol{x}) \\ 
\psi ^{\text{f}\downarrow }(\boldsymbol{x})%
\end{pmatrix}%
& \longrightarrow e^{i\alpha (\boldsymbol{x})}%
\begin{pmatrix}
\psi ^{\text{f}\uparrow }(\boldsymbol{x}) \\ 
\psi ^{\text{f}\downarrow }(\boldsymbol{x})%
\end{pmatrix}%
,  \notag \\
\begin{pmatrix}
\psi ^{\text{b}\uparrow }(\boldsymbol{x}) \\ 
\psi ^{\text{b}\downarrow }(\boldsymbol{x})%
\end{pmatrix}%
& \longrightarrow e^{i\beta (\boldsymbol{x})}%
\begin{pmatrix}
\psi ^{\text{b}\uparrow }(\boldsymbol{x}) \\ 
\psi ^{\text{b}\downarrow }(\boldsymbol{x})%
\end{pmatrix}%
.  \label{BLocalU2}
\end{align}%
Corresponding to these two U(1) symmetries, we should introduced two
normalized CB fields by%
\begin{equation}
\phi ^{\text{f}\alpha }(\boldsymbol{x})=\sqrt{\rho ^{\text{f}}(\boldsymbol{x})}n^{%
\text{f}\alpha }(\boldsymbol{x}),\quad \phi ^{\text{b}\alpha }(\boldsymbol{x})=\sqrt{%
\rho ^{\text{b}}(\boldsymbol{x})}n^{\text{b}\alpha }(\boldsymbol{x}),  \label{TwoCP1}
\end{equation}%
where $\alpha =\uparrow ,\downarrow $ and 
\begin{equation}
\sum_{\alpha =\uparrow \downarrow }n^{\text{f}\alpha \dagger }(\boldsymbol{x})n^{%
\text{f}\alpha }(\boldsymbol{x})=\sum_{\alpha =\uparrow \downarrow }n^{\text{b}%
\alpha \dagger }(\boldsymbol{x})n^{\text{b}\alpha }(\boldsymbol{x})=1.
\end{equation}%
We have a set of two CP$^{1}$ fields as the basic fields, each of which is
the dynamic field for each layer at $\nu =2$. Topological solitons are
simple skyrmions.

It is interesting to consider a case without the interlayer exchange
interaction ($J^{d}\simeq 0$) but with a nonnegligible tunneling interaction
(${\Delta }_{\text{SAS}}\not=0$). The basic field is the CP$^{3}$ field
because the Hamiltonian possesses only the local U(1) symmetry (\ref%
{B2LocalU1}). Hence, we have the G$_{4,2}$ field at $\nu =2$. It is a
genuine BLQH system and topological solitons are biskyrmions.

\begin{figure}[tbph]
\includegraphics*[width=75mm]{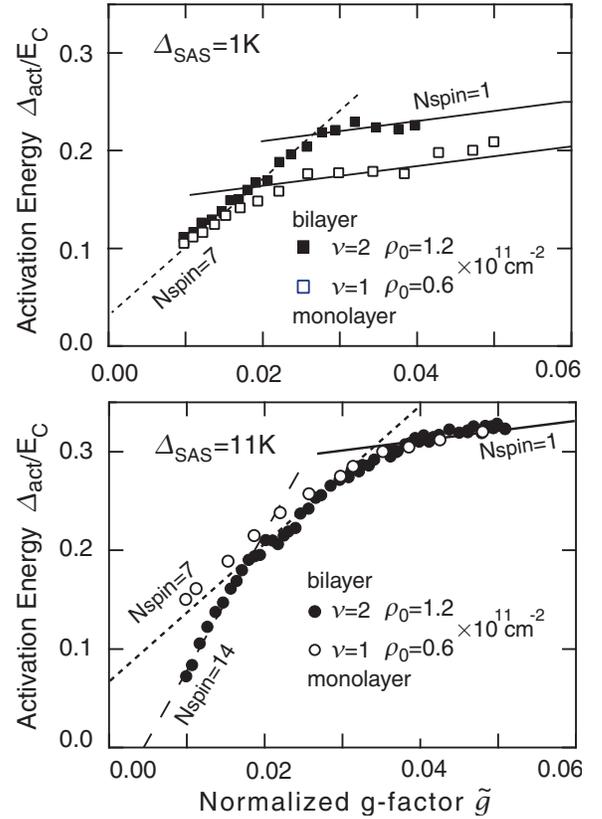}
\caption{ The activation energy is given in the $\protect\nu =2$ BLQH state
(solid marks) and the $\protect\nu =1$ monolayer QH state (open marks). The
data are taken from Kumada et al.\protect\cite{Kumada00L}. The vertical axis
is the activation energy $\Delta _{\text{act}}$ in units of $E_{\text{C}%
}^{0}\equiv e^{2}/4\protect\pi \protect\varepsilon \ell _{B}$. The
horizontal axis is the normalized $g$-factor $\widetilde{g}\equiv {\Delta }_{%
\text{Z}}/E_{\text{C}}^{0}$. The number of flipped spins is given by the
slope of the data $\text{N}_{\text{spin}}=\partial \Delta _{\text{act}%
}/\partial \widetilde{g}$. }
\label{ExperKumad}
\end{figure}

We wish to do a thinking experiment to make it convincing that a biskyrmion
is excited even for $J^{d}\simeq 0$ provided ${\Delta }_{\text{SAS}}\neq 0$.
We start with the SU(4)-invariant limit of the exchange interaction ($%
J^{d}\simeq J$), where a biskymion must be excited. We question what would
happen as $J^{d}$ is decreased. As far as ${\Delta }_{\text{SAS}}$ is
significant, two electrons in one Landau site are indistinguishable, and
hence we need the Grassmannian field $Z(\boldsymbol{x})$ to describe the system.
Furthermore, the spin biskyrmion (\ref{SpinSkyrm}) is insensible to the
interlayer stiffness $J^{d}$, as we have remarked, because it is governed by
the Hamiltonian (\ref{ContiSpin}). Hence, nothing would happen for the spin
biskyrmion as $J^{d}\rightarrow 0$. It is to be stressed that the existence
of topological solitons is the property of the Grassmannian manifold and not
the property of the Hamiltonian (\ref{ExchaGrass}). When $J^{d}\simeq 0$ the
Grassmannian soliton is not the BPS state of the Hamiltonian, but its very
existence is guaranteed by the homotopy theorem (\ref{SpinBreak}).

Topological solitons arise as quasiparticles (charged excitations). Their
excitation energy is observed as the activation energy. By measuring it as a
function of the tilting angle $\Theta $, we can tell how many spins are
flipped by one topological soliton\cite{Schmeller95L}. Kumada et al.\cite%
{Kumada00L} made a careful measurement of activation energy by using two
bilayer samples with a large tunneling gap (${\Delta }_{\text{SAS}}\simeq 11$
K) and a small tunneling gap (${\Delta }_{\text{SAS}}\simeq 1$ K). These two
samples have precisely the same sample parameters except for the tunneling
gap, where $J^{d}/J=0.15$ with use of (\ref{BPpinStiff}). Thus, $J^{d}$ is
quite small compared with $J$. They have also measured activation energy in
the monolayer limit of the same samples. (The monolayer state is constructed
by emptying the back layer by tuning the bias voltage in the bilayer sample.
The total electron density in the bilayer system is controlled so that it is
precisely twice of that in the monolayer system.) They have found 7 flipped
spins in the 1K-sample while 14 flipped spins in the 11K-sample when the
tilting angle is small [Fig.\ref{ExperKumad}]. When the tilting angle
becomes large, the number of flipped spins makes a transition from 14 to 7
in the 11K-sample. This is understood as follows. As the sample is tilted,
the tunneling gap is known\cite{Hu92B} to decrease as%
\begin{equation}
{\Delta }_{\text{SAS}}(\Theta )={\Delta }_{\text{SAS}}\exp \{-(d/2\ell
_{B})^{2}\tan ^{2}\Theta \}.
\end{equation}%
In Fig.\ref{ExperKumad} the transition occurs at $\Theta =60^{\circ }$,
where ${\Delta }_{\text{SAS}}\simeq 2$K. It is small enough compared with
other interactions and would be practically negligible. They have also
confirmed 7 flipped spins in the monolayer limit of both samples. (Only one
spin is flipped in all cases for a sufficiently large tilting angle, where a
vortex is excited in one of the layers since the tunneling gap becomes so
small and the Zeeman effect becomes so large.) These facts are consistent
with our conclusion that biskyrmions (simple skyrmion) are excited in a
sample with a large (negligible) tunneling gap.

\section{Summary}

\label{SecSumma}

We have investigated the BLQH systems at $\nu =2$. There are three phases,
i.e. the spin phase, the ppin phase and the canted phase. Experimentally the
spin and ppin phases are clearly observed. We have presented a
field-theoretical formulation of these two phases, and analyzed soft waves
and topological excitations. We have shown that the dynamic field is the
Grassmannian G$_{4,2}$ field, which is a set of two CP$^{3}$ fields but
contains only four complex independent fields. Accordingly there are four
independent soft waves (pseudo-Goldstone modes) as neutral low energy
excitations. We have also shown that there are topological excitations as
charged excitations: They are biskyrmions comprised of two simple skyrmions
in the two layers and possess the electric charge $2e$. This conclusion on
charged excitations is confirmed by a recent experimental result\cite%
{Kumada00L} in the spin phase at $\nu =2$, where the number of flipped spins
is found to be twice as large as that at $\nu =1$ for a sample with large ${%
\Delta }_{\text{SAS}}$.

\section*{Acknowledgement}

We would like to thank to Anju Sawada, Norio Kumada and Hiroshi Ishikawa for
fruitful discussions. A part of this work was done when one of the authors
(ZFE) was a visiting professor at Institute of Solid State Physics, Tokyo
University. He is very grateful to Yasuhiro Iye and Shingo Katumoto for
their warm hospitality.

\end{document}